\documentclass{aa}  

\usepackage{graphicx}
\usepackage{color}
\usepackage{amsmath}
\usepackage{array}
\usepackage{listings}
\usepackage{txfonts}
\begin{document}

   \title{Broadening of the DEM by multi-shelled and turbulent loops}

   %\subtitle{I. Overviewing the $\kappa$-mechanism}

   \author{T. Van Doorsselaere
          \inst{1}
          \and
          P. Antolin\inst{2,3}
          \and 
          K. Karampelas\inst{1}
          }

   \institute{Centre for mathematical Plasma Asytrophysics, 
   Mathematics Department, KU~Leuven,\\
   Celestijnenlaan 200B bus 2400, 3001 Leuven, Belgium\\
              \email{tom.vandoorsselaere@kuleuven.be}
         \and
             {School of Mathematics and Statistics, University of St. Andrews, St. Andrews, Fife KY16 9SS, UK} 
         \and {National Astronomical Observatory of Japan, Osawa, Mitaka, Tokyo 181-8588, Japan}
             }

   \date{Received xx; accepted xx}

% \abstract{}{}{}{}{} 
% 5 {} token are mandatory
 
  \abstract
   {Broad differential emission measure (DEM) distributions in the corona are a sign of multi-thermal plasma along the line-of-sight. Traditionally, this is interpreted as evidence of multi-stranded loops. Recently, however, it has been shown that multi-stranded loops are unlikely to exist in the solar corona, because of their instability to transverse perturbations.}
   {We aim to test if loop models subject to the Transverse Wave-Induced Kelvin-Helmholtz (TWIKH) instability result in broad DEMs, potentially explaining the observations.}
   {We took simulation snapshots and compute the numerical DEM. Moreover, we performed forward-modelling in the relevant AIA channels before reconstructing the DEM. }
   {We find that turbulent loop models broaden their initial DEM, because of the turbulent mixing. The width of the DEM is determined by the initial temperature contrast with the exterior.}
   {We conclude that impulsively excited loop models have a rather narrow DEM, but that continuously driven models result in broad DEMs that are comparable to the observations.}


   \maketitle
%
%________________________________________________________________

\section{Introduction}
Differential emission measure (DEM) inversion is a technique to use multi-band photometry or spectrometry and convert it to a temperature distribution of the plasma along the line-of-sight \citep[see, e.g.][]{cheung2015}. Given the ubiquity of observations from SDO/AIA, several independent methods have been developed to invert the emission in the different AIA filters to a DEM distribution \citep{aschwanden2011c,guennou2012,plowman2013,cheung2015}. \citet{testa2012} used the method by \citet{kashyap1998} to show that the DEM inversions from the AIA filters have lower accuracy in the high temperature range \citep[and the same was shown for the combination of  XRT and Hinode/EIS,][]{winebarger2012}.\par

The DEM inversion method is often used to observationally study the thermal structure of coronal loops \citep[e.g.][]{reale2009,schmelz2011}. Often, it is found that coronal loops are multi-thermal, in the sense that they display a broad temperature distribution in the differential emission measure \citep{schmelz2014}. Typically, this is interpreted as the proof of the presence of multi-stranded (or multi-threaded) loops aligned with the magnetic field \citep[e.g.][]{aschwanden2000,klimchuk2015}, and this is supported by forward modelling as well \citep{brooks2012,viall2015} and by high-resolution observations \citep{peter2013}.\par  

In multi-stranded loop models, each of the strands has a thermodynamic evolution independent of the other strands. In this approximation, the magnetic field is assumed to play no role in the longitudinal, thermodynamic evolution of these strands. Additionally, the thermal conduction perpendicular to the magnetic field is taken to be small and the structures are considered to be optically thin (and are thus not absorbing the radiation from the neighbouring strands). With these assumptions, there is no mechanism for energy exchange between the multiple strands. In such models, each strand is heated with nano-flares \citep{viall2013,klimchuk2015}, usually assumed to be because of small-scale reconnection between different threads. \par

Since a few years, we know that the solar corona is filled with transverse \citep{tomczyk2007,nistico2013} and longitudinal waves \citep{krishnaprasad2012}. Numerical simulations show that these ubiquitous transverse oscillations have a strong effect on coronal loop models, because the transverse waves trigger the Kelvin-Helmholtz instability (KHI) in the loop boundaries \citep{terradas2008b}, resulting in a turbulent regime. This instability has previously been simulated in different conditions corresponding to photospheric and chromospheric structures \citep{Karpen_1993ApJ...403..769K,Ofman_1994GeoRL..21.2259O,Poedts_1997SoPh..172...45P, Ziegler_1997AA...327..854Z}. In a loop configuration, the multi-shelled structure at different radii from the centre present an Alfv\'en continuum due to their continuous change in density, which is expected to generate quasi-modes from resonant absorption \citep[e.g.][]{goossens1992} and mode coupling \citep{DeMoortel_2016PPCF...58a4001D}. This means that in such multi-shelled loops there will be a continuous transfer of energy from global oscillation modes (the kink wave) into these shells. This energy is manifested, in particular, as an increase in the azimuthal velocity, that is, the velocity along these shells, and ends up in the turbulence after the KHI sets in. In these models there is therefore a strong redistribution and mixing of the energy and plasma in the transverse direction. \par

\citet{antolin2014} show that such KHIs are expected at basically any amplitude for impulsive transverse velocity perturbations of long coronal loops (for which the radius-to-length ratio is small), and used forward modelling to show that the transverse wave-induced Kelvin-Helmholtz (TWIKH) roll-ups may make the loop appear as multi-stranded in high-resolution observations. \citet{karampelas2018} model coronal loops that are driven by transverse waves at the footpoint, and showed that the loop becomes entirely turbulent, resulting in a strong mixing of all of the loop plasma with the exterior. 

It had been suggested that finite twist or transport coefficients may inhibit the development of the KHI. The effect of twist on the TWIKH rolls is investigated by \citet{howson2017b,terradas2018} and the effect of viscosity and resistivity by \citet{howson2017}. All these studies show that the onset of the KHI may be delayed because of these effects, but the KHI is not stabilised.

Considering the effect of transverse waves on multi-stranded loop models, \citet{magyar2016} showed that an initially multi-stranded loop is thoroughly mixed by the TWIKH rolls. Such a mixing of plasma is not at all considered in the multi-stranded loop models. The mixing results also in a turbulent state of the loop plasma, which can no longer be considered as multi-stranded in the conventional sense. Thus, these recent simulations show that the multi-stranded loops are not stable to the omnipresent transverse motions, and thus it is unlikely that such loops exist \citep[given the omnipresence of transverse oscillations,][]{anfinogentov2015}. Therefore, the question arises if other models than multi-stranded loops (in the conventional sense, i.e., composed of independent strands) can explain the observed broad DEMs. Here we will study the influence of the TWIKH rolls on the DEM, and we shall investigate if the loop models with TWIKH rolls can explain the broad DEMs as well, given that the TWIKH rolls keep similar filling factors and emission measures \citep{magyar2016}.\\ 
In order to perform this investigation, we first explain the numerical methods that are used in Sect.~\ref{sec:methods}. Then, we consider DEMs of multi-shelled loops (Sect.~\ref{sec:shell}), before constructing DEMs for impulsively excited loops (Sect.~\ref{sec:standing}) and driven loops (Sect.~\ref{sec:driven}). We discuss the implications of our results in Sect.~\ref{sec:conclusions}.

\section{Methods}
\label{sec:methods}
In this paper, several methods have been used to construct the DEM of numerical data. 
\subsection{Numerical DEM}\label{sec:numerical}
In general, the DEM is defined by the equation
\begin{equation}
        \mbox{EM}= \int n_\mathrm{e}^2 dz = \int n_\mathrm{e}^2 \frac{dz}{dT} dT \equiv \int \mbox{DEM}(T) dT,
\end{equation}
where EM is the emission measure, $n_\mathrm{e}$ is the electron number density, and $z$ is the coordinate along the line-of-sight. To numerically construct the DEM in a simulation for a particular pixel, we extract from the simulation the distribution of the density $n_\mathrm{e}(z)$ (array \verb+ne+) and temperature $T(z)$ (array \verb+te+) along the line-of-sight, with a similar method to \citet{peter2006}. We used the following IDL code to construct the DEM:
\begin{lstlisting}[language=idl]
 nbins=33
 deltax=(x_grid[1]-x_grid[0]) ; in cm!
 dem=make_array(nbins)
 ; te in K!
 thist=histogram(alog10(te[*]),nbins=nbins, $
        locations=locs,reverse_indices=R, $
        min=5.5,max=6.8)
 deltalocs=locs[1]-locs[0]
 for i=0,nbins-1 do begin
 ; ne in cm^{-3}!
     if R[i] ne R[i+1] then dem[i]+= $
         total(ne[R[R[i] : R[i+1]-1]]^2)* $
         deltax/10^locs[i]/(10^deltalocs-1) 
 endfor  
\end{lstlisting}
Thus, the DEM is constructed as a histogram of $\log{T}$ with 33 bins between $\log{T_\mathrm{min}}=5.5$ and $\log{T_\mathrm{max}}=6.8$, with weights of $n_\mathrm{e}^2$. In essence, the DEM is the probability density function (PDF) of the electron density squared ($n_\mathrm{e}^2$) as a function of $\log{T}$, multiplied by the total emission measure (EM). In what follows, we only construct DEMs using `infinite' resolution, meaning that we take the values on a strip along the line-of-sight with a width of only one simulation pixel. We have investigated the effect of including macro-pixels by considering wider strips, but its effect is only to smooth the DEM distributions: they show fewer gaps, and are smoother functions in that case. \\
Following the same procedure for a location outside of the loop, we also compute the DEM of the background. Then we subtracted the background DEM from the relevant DEM to mimic background subtraction. In what follows, we always show background subtracted DEMs only. 

% After the DEM construction for a single pixel, we average that DEM array with those for the 10 neighbouring pixels, in order to simulate a finite resolution of the image data. 

%\subsubsection{Characterisation}
When displaying the obtained numerical $\log{\mbox{DEM}}$ as a function of $\log{T}$, the typical shape is a trapezoid-like, bound between $\log{T_\mathrm{min}}$ and $\log{T_\mathrm{max}}$ (see middle panel of Fig.~\ref{fig:layerwidth}), and a peak at the internal loop temperature. In between, the $\log{\mbox{DEM}}$ has a curved shape, with a tail at the background value of the temperature of 1MK.\\ Instead of using the standard expression for the standard deviation and the full-width-half-maximum (FWHM), we characterise the width of the distribution by computing 
\begin{equation}\sigma=\sqrt{\frac{\sum_{T>1.1MK} n_\mathrm{e}^2 (\log{T}-\log{T_\mathrm{max}})^2 }{\sum_{T>1.1MK} n_\mathrm{e}^2}},\end{equation}
so we define $\sigma$ as the standard deviation of the temperature distribution (for $T>1.1$MK) from the theoretical value of $T_\mathrm{max}$, with $n_\mathrm{e}^2$ as weights. The reason for not using the standard expressions for the standard deviation is that the mean temperature is decreasing in the dynamic models, especially in the later part of the simulations when the loops become turbulent (see later). The above expression for the width of these asymmetric distributions is less influenced by temperature changes, and allows to disentangle the effects of temperature change and broadening. The obtained value for the width of the distribution is then converted to the FWHM with the formula
\begin{equation}\text{FWHM}_\mathrm{numerical}=\sigma \sqrt{2\ln{2}},\end{equation}
where we have taken into account that it is only half the width of a normal Gaussian, because of the special asymmetric shape of the DEM.

\subsection{DEM by forward modelling}
We also used an alternative method to first construct a forward model of the simulation snapshots and afterwards perform a DEM inversion. Each model is processed with version 3.3 of the FoMo code\footnote{The FoMo code can be downloaded from \url{https://wiki.esat.kuleuven.be/FoMo/}} \citep[as described in][]{vd2016fomo}. The code computes the emission of the model in the six filters (94\AA{}, 131\AA{}, 171\AA{}, 193\AA{}, 211\AA{}, 335\AA{}) of SDO/AIA \citep{boerner2012}, using the method described in \citet{delzanna2011,yuan2015}. For the computation of the filter intensities, we used the \texttt{sun\_coronal.abund} file of CHIANTI v8 \citep{dere1997,landi2013}.\par
Then, a DEM inversion is performed on the emission of the forward model. Before performing the DEM analysis, we performed a background subtraction: from the emission in the loop, we subtract the emission observed at $r=2.41\mathrm{Mm}$ (which is well outside the loop region). This is a standard procedure when performing DEM analysis of coronal loop observations \citep[e.g.][]{tripathi2009}. For the DEM inversion, we used the regularised inversion code which was published by \citet{hannah2012}. As error estimates for the emission, we put a nominal value of 10\% of the pixel's emission. We have performed the DEM inversion in the temperature interval $\log{(T)}\in[5.5,6.8]$, with 33 temperature bins, as we also did in the DEM without forward modelling.\par
We characterised the forward modelled DEM by computing the full width half maximum (FWHM), assuming that the DEM distributions are close to log-normal \citep[e.g.][]{cheung2015}. These parameters are measured by fitting a parabola to the logarithm of the DEM:
\begin{equation}
        \log{(\mbox{DEM})} = a \log^2{(T)} + b \log{(T)} + c,
        \label{eq:parabola}
\end{equation}
with $a, b$, and $c$ the fitting parameters. The position of the maximum $T_\mathrm{max,FoMo}$ and the FWHM of the DEM distribution is calculated by
\begin{equation}
        \log{(T_\mathrm{max,FoMo})}=-\frac{b}{2a}, \qquad 
        \mbox{FWHM}_\mathrm{FoMo}=2\sqrt{-\frac{\ln{2}}{a}}.
        \label{eq:fitting}
\end{equation}
For the fitting of the profile, we took a range of temperatures around $T_\mathrm{max}$ with a width of 0.32 (in $\log{T}$). We only considered the positive DEM values.

%\subsection{Models}
%\label{sec:models}
%We have selected several analytical and numerical models for coronal loops. 

\section{Results}
\label{sec:results}

\subsection{DEM for multi-shelled loops}\label{sec:shell}
\begin{figure}
        \includegraphics[width=\linewidth]{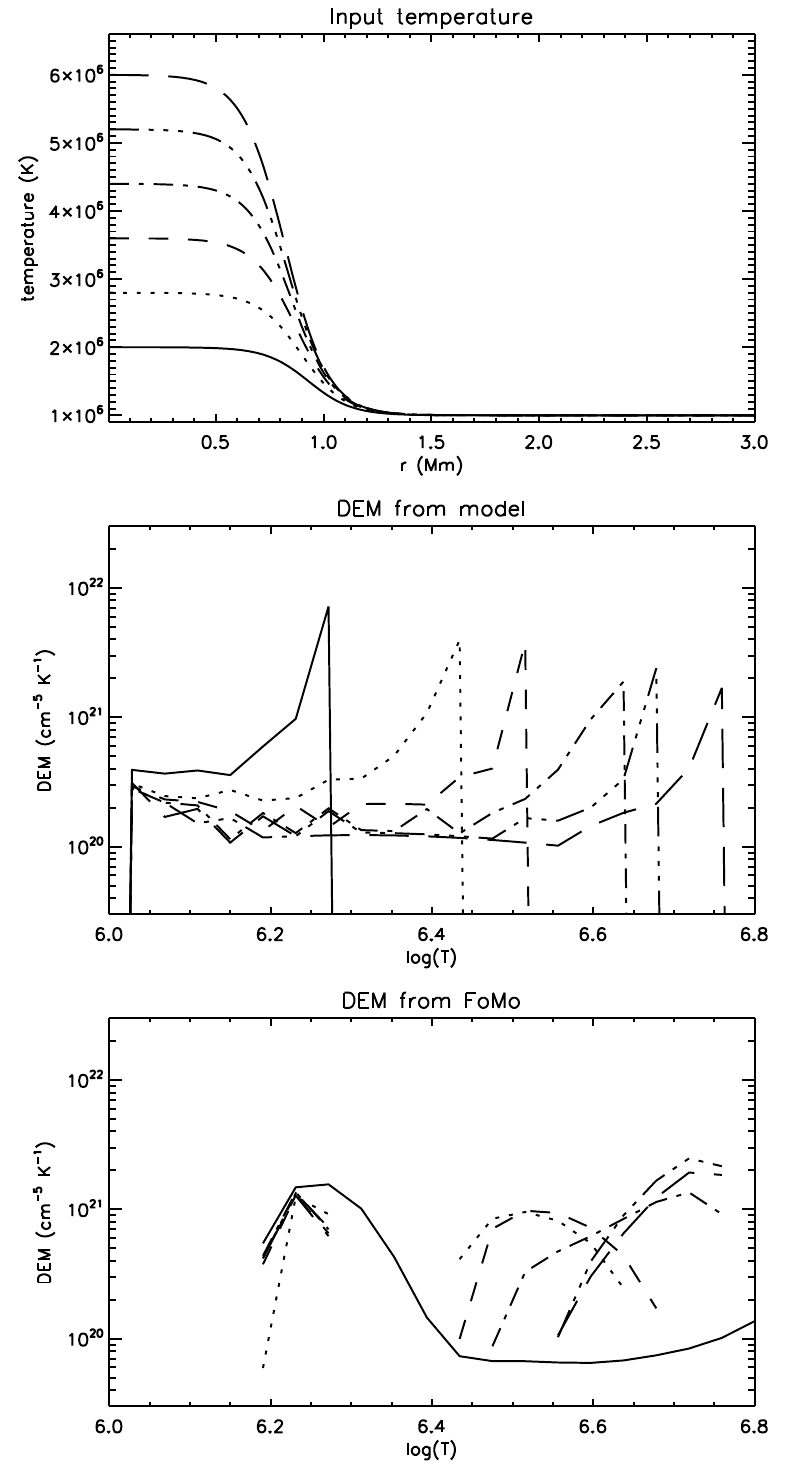}
        \caption{Top panel: Different temperature distributions as a function of radius. The linestyle in all panels corresponds to the temperature profile given with that linestyle in the top panel. Middle panel: Numerical DEM without forward modelling for the LOS through the centre of the loop. Bottom panel: DEM after forward modelling and DEM inversion. }
        \label{fig:layerwidth}
\end{figure}

First we constructed simple 1D cylindrical models for a loop, with only variation of the density and temperature in the radial direction. The radial variation of the density and temperature can be considered as multi-shelled. For the density profile, we take 
\begin{equation}
        n_\mathrm{e}(r)=n_\mathrm{e,min}+\frac{(n_\mathrm{e,max}-n_\mathrm{e,min})}{2}\left\lbrace 1-\tanh{\left(d\left(\frac{r}{R}-1\right)\right)}\right\rbrace,\label{eq:density}
\end{equation}
with the parameters 
\[n_\mathrm{e,min}=10^9 \mbox{cm}^{-3}, \quad n_\mathrm{e,max}=3\ 10^9 \mbox{cm}^{-3}, \quad R=1\mbox{Mm}, \quad d=5.\] This results in a loop that is three times as dense as the surroundings. For the temperature profile, we similarly take 
\begin{equation}
        \frac{1}{T(r)}=\frac{1}{T_\mathrm{min}}+\left(\frac{1}{T_\mathrm{max}}-\frac{1}{T_\mathrm{min}}\right)\frac{1}{2} \left\lbrace 1-\tanh{\left(d\left(\frac{r}{R}-1\right)\right)}\right\rbrace,\label{eq:temperature}
\end{equation}
with $T_\mathrm{min}=1$MK. For the central, peak temperature, we consider six values: $\lbrace 2,2.8,3.6,4.4,5.2,6\rbrace$ MK. Each of the temperature profiles is drawn in the top  panel of Fig.~\ref{fig:layerwidth}, each different linestyle corresponding to a different peak temperature. 

\begin{figure*}
        \includegraphics[width=\linewidth]{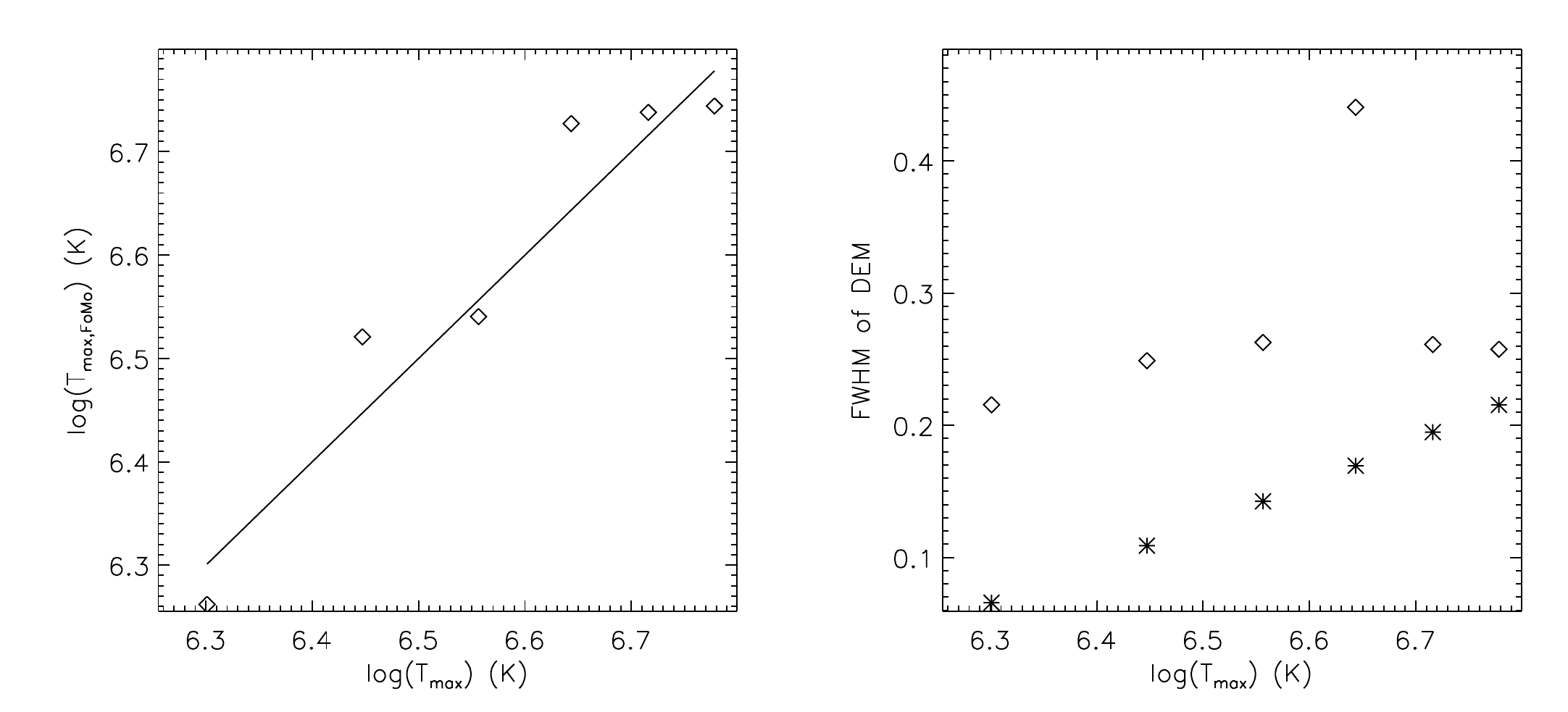}
        \begin{center} \begin{tabular}{llll}
                \hline\hline
                $\log{(T_\mathrm{max})}$ (K) & $\log{(T_\mathrm{max,FoMo})}$ (K) & FWHM$_\mathrm{numerical}$ ($\log{T}$) & FWHM$_\mathrm{FoMo}$ ($\log{T}$)\\
\hline
6.301 & 6.261 & 0.074 & 0.214\\
6.447 & 6.520 & 0.124 & 0.253\\
6.556 & 6.542 & 0.163 & 0.262\\
6.643 & 6.716 & 0.195 & 0.405\\
6.716 & 6.736 & 0.226 & 0.260\\
6.778 & 6.744 & 0.251 & 0.259\\
\hline
        \end{tabular}\end{center}
        \caption{Left panel: Peak temperature of the DEM from the forward model compared to the input $T_\mathrm{max}$. Right panel: FWHM of the DEM as a function of the peak temperature of the profile ($T_\mathrm{max}=T(r=0)$), for the numerical DEM without forward modelling ($\ast$, middle panel of Fig.~\ref{fig:layerwidth}) and the forward modelled DEM ($\Diamond$, bottom panel of Fig.~\ref{fig:layerwidth}). Bottom table: Numerical values for the peak temperature and widths of the DEM.}
        \label{fig:peakfwhm}
\end{figure*}

For each of the temperature profiles, we constructed the numerical DEM and the DEM by forward modelling. These are shown by the corresponding linestyles in the middle and bottom panels of Fig.~\ref{fig:layerwidth}, respectively. The first thing to see in Fig.~\ref{fig:layerwidth} is that the DEMs from both models do not agree very well. The DEM from the forward modelling is much broader than the DEM without the forward modelling, which are printed for clarity in the table in Fig.~\ref{fig:peakfwhm}. Still, the peak temperature of the DEM from the forward modelling corresponds rather well with the input maximum temperature $T_\mathrm{max}$, as is shown in the left panel of Fig.~\ref{fig:peakfwhm}. Thus, from this limited parameter study, we can conclude that the DEM in our artificial observations measures the position of the peak temperature well. On the other hand, the DEM from the forward modelling does not capture the true width of the DEM. This is displayed in the right panel of Fig.~\ref{fig:peakfwhm}. The width of the theoretical DEM is shown with stars, and the width of the DEM from forward modelling is shown with diamonds. It is clear that the DEM from the forward modelling is most of the time much broader than the numerical DEM. \par

For the later results, it is important to notice the behaviour of the FWHM of the numerical DEM (without forward modelling) with the peak temperature. The higher the peak temperature, the broader the distribution, and these two quantities show a strong correlation (see right panel of Fig.~\ref{fig:peakfwhm}). This is also readily visible in the middle panel of Fig.~\ref{fig:layerwidth}. This is, however, not true for the DEM from the forward modelling. Neglecting the very high value for $\log{T_\mathrm{max}}=6.72$, the width of the DEM from the forward modelling seems to be nearly insensitive to the input value of $T_\mathrm{max}$.

\subsection{DEM for impulsively excited loops}\label{sec:standing}

The next model to consider is a dynamic evolution of a loop with a sharp boundary, based on the simulations performed in \citet{antolin2014,antolin2015}, using the CIP-MOCCT code \citep{kudoh1999}. From a parameter study, we estimate that the effective (combined explicit and numerical) Reynolds and Lundquist numbers in the code are of the order of $10^4-10^5$, with some anomalous resistivity included to stabilise the simulation. 

The density and temperature in the initial state are taken to have a discontinuous step, corresponding to $d\to\infty$. The density parameters are as before, and the temperature ranges from 1MK (outside) to 3MK (inside). The magnetic field is adjusted in order to keep the perpendicular pressure balance. \par

\begin{figure*}
        \includegraphics[width=\linewidth]{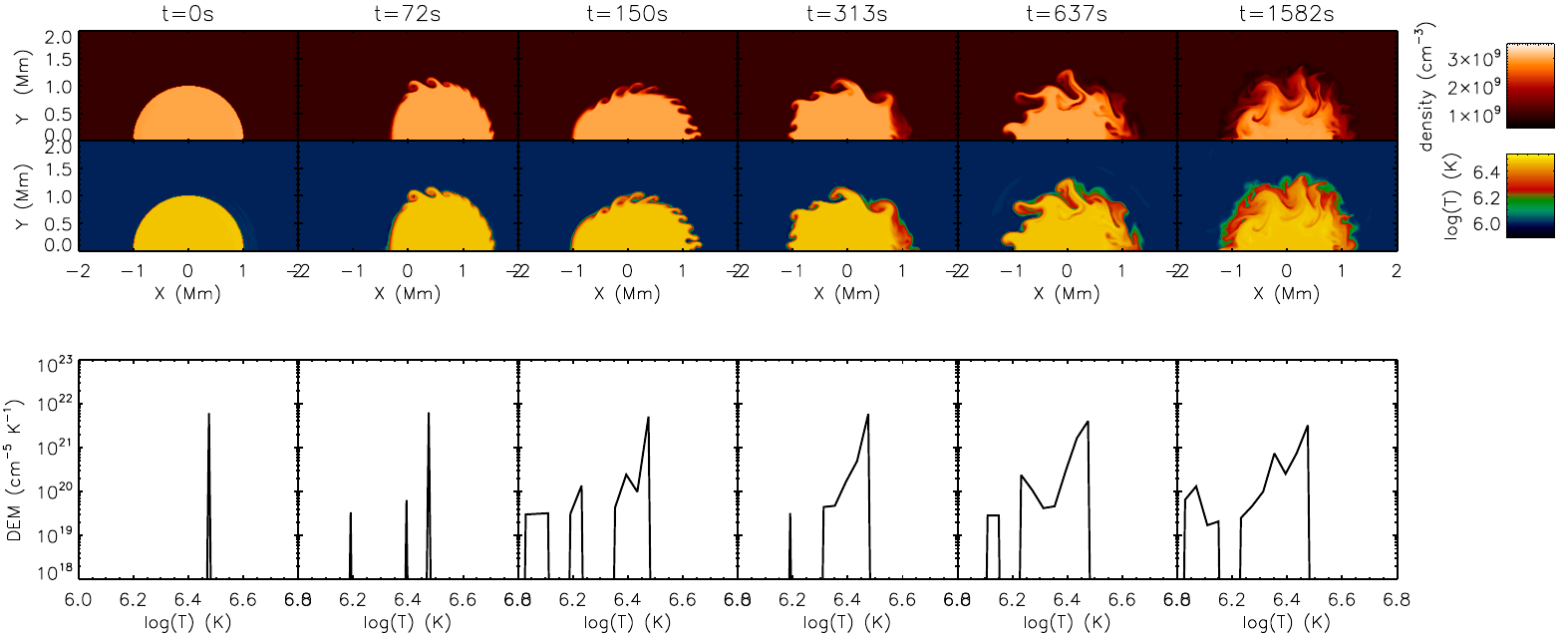}
        \caption{Evolution of density (top row), temperature (middle row) and DEM (bottom row) in the loop top cross section of the impulsively excited model, at the times specified in the title of each column. The DEM was computed for a single pixel corresponding to the LOS through the centre-of-mass, perpendicular to the initial motion.}
        \label{fig:demovertime}
\end{figure*}

The static loop model is perturbed by a velocity field at the start of the simulation, with dependence
\begin{equation}
        \vec{v} = v_0 \cos(\pi z/L)\zeta(r)\vec{e}_x,
\end{equation}
where $\vec{e}_x$ is the unit vector in the $x$-direction, $v_0=0.05~c_s$ is the initial amplitude, with $c_s$ the sound speed. This corresponds to $v_0=$15~km~s$^{-1}$ in this model. $\zeta(r)$ is the Heaviside function: the driver is finite within the loop, and zero outside the loop. As a result, the loop starts to oscillate transversally with a kink mode, at its fundamental harmonic with a period of $P=315$s. The kink mode experiences large shear motions near the loop boundary, and these lead to the formation of TWIKH roll-ups at the edge of the coronal loop \citep[as was first shown by][]{terradas2008b}, also known as TWIKH rolls. Then, the Kelvin-Helmholtz roll-ups further cascade to form a turbulent regime, smearing out the density and temperature profile radially \citep{magyar2016}. The temporal evolution of the density and temperature is shown in the top two rows of Fig.~\ref{fig:demovertime}, where the shown times correspond to $t=0$, $t=P/4$, $t=P/2$, $t=P$, $t=2P$ and $t=5P$. For more details on the models, we advise the readers to consult \citet{antolin2014,antolin2016}. 

In the bottom row, the numerical DEM is shown for the corresponding time of the evolution. The DEM is computed for the line-of-sight through the centre-of-mass of the loop perpendicular to the original perturbation, and the DEM of the background has been subtracted as before. In the left panel (for $t=0$), the DEM shows a sharp peak at 3MK, as expected in the original configuration. Over time, the peak spreads more to the lower (exterior) temperatures, gradually filling up the entire range between the exterior temperature of 1MK and the internal temperature of 3MK. This is expected because of the mixing effect of the TWIKH rolls, mixing the interior, hot plasma, with the exterior, cool plasma \citep{magyar2016,karampelas2017}. The resulting DEM is thus compatible with our conclusions of Sect.~\ref{sec:shell}: the turbulent density and temperature from the TWIKH rolls evolve into a smooth boundary layer. In that smooth boundary layer, the emission follows the same behaviour as the static models in Sect.~\ref{sec:shell}, resulting in a `trapezoid' shaped DEM. 

\begin{figure}
        \includegraphics[width=\linewidth]{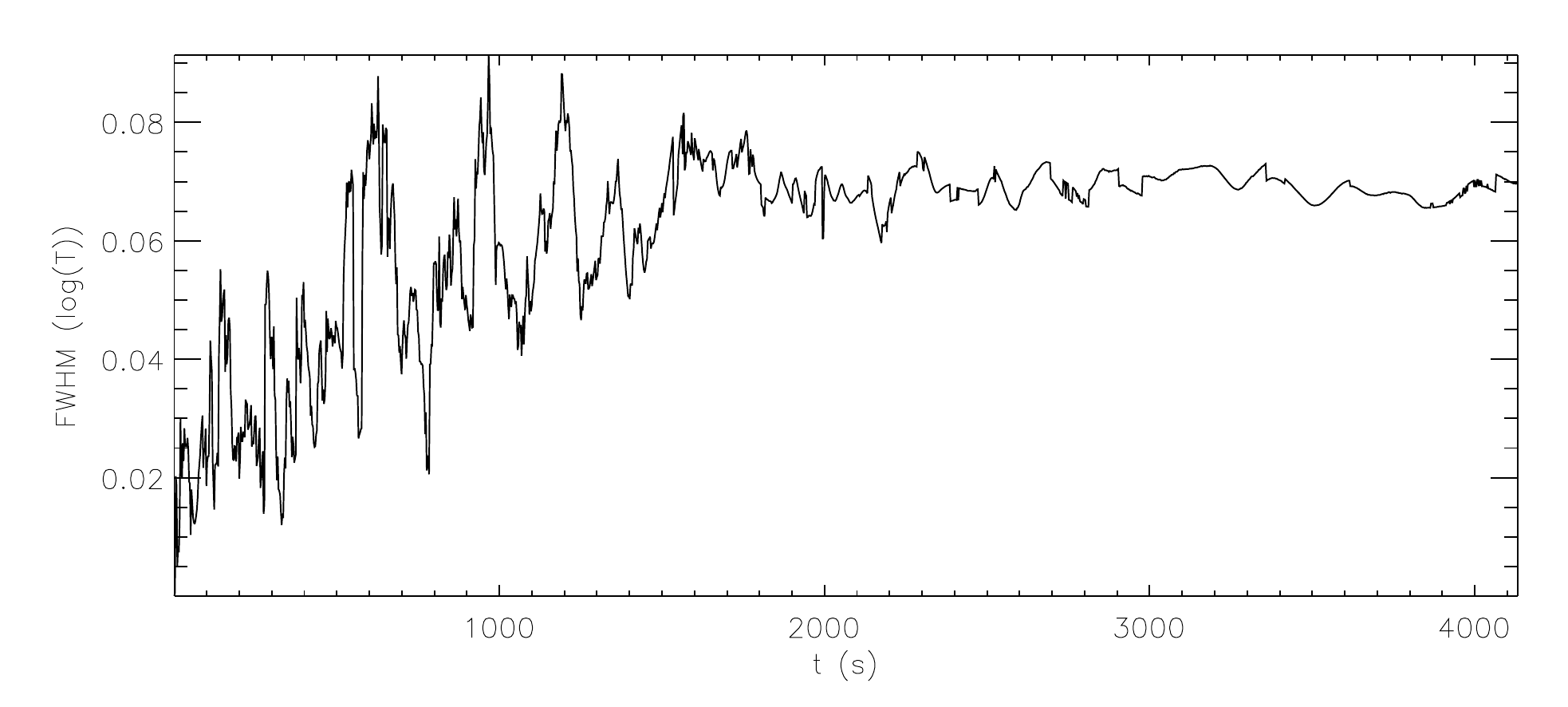}
        \caption{FWHM of the numerical DEM (without forward modelling) as a function of time, for the impulsively excited model.}
        \label{fig:fwhm}
\end{figure}

We compute the FWHM of the DEM, following the method outlined in Eq.~\ref{eq:parabola}-\ref{eq:fitting}. The evolution of the FWHM of the DEM as a function of time is shown in Fig.~\ref{fig:fwhm}. The figure confirms our understanding of the DEMs in the bottom row of Fig.~\ref{fig:demovertime}. Initially, there is a very sharp DEM, and this is confirmed with a FWHM of around 0. Then, the loop plasma is mixed with the exterior because of the TWIKH rolls, and this results in a (roughly linear) increase in the FWHM over the first four periods of the oscillation. This is the first important conclusion in this paper: {the DEM of loops is broadened by the presence of transverse oscillations.} Then, the FWHM settles to fluctuate around a constant final value of around 0.07 (in $\log(T)$, corresponding to 17\%). This value is somewhat lower than what can be inferred from the right panel of Fig.~\ref{fig:peakfwhm}, but this discrepancy can be attributed to having a larger $d$, which would decrease the FWHM of the DEM. The saturation of the FWHM is to be expected, because the oscillation was impulsively excited. At the time the FWHM saturates, the global oscillation is nearly completely damped and the TWIKH rolls are no longer amplified. As we understand from the previous Sect.~\ref{sec:shell}, the final value of the FWHM is mainly determined by the initial temperature profile, and particularly the temperature contrast between the interior and exterior. 

\subsection{DEM of driven loops}\label{sec:driven}
\begin{figure*}
        \includegraphics[width=\linewidth]{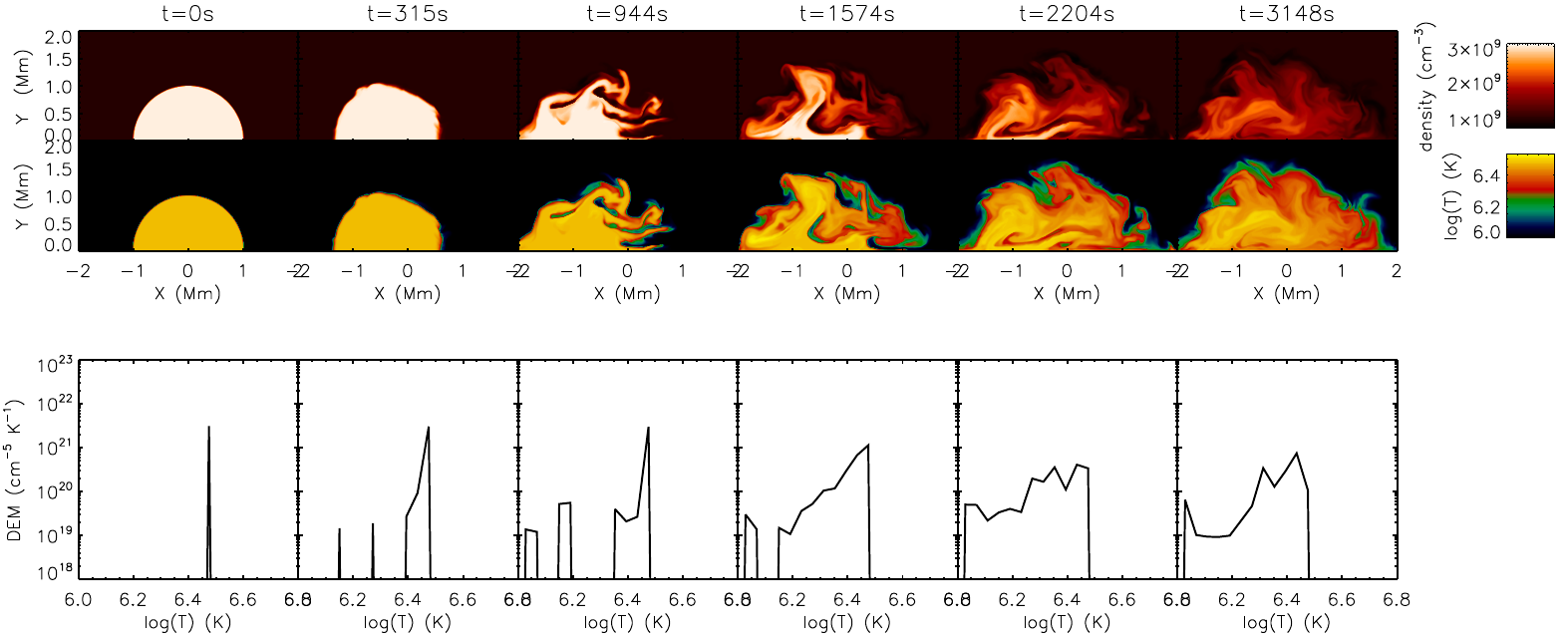}
        \caption{Same as Fig.~\ref{fig:demovertime}, but for the driven model (Sect.~\ref{sec:driven}). }
        \label{fig:demovertime_kostas}
\end{figure*}

In our final model, we consider the models of \citet{karampelas2018}, which have been extended for a longer time from \citet{karampelas2017}. These models are the driven counterpart of the models by \citet{antolin2014}: the excitation is no longer by an initial velocity pulse, but rather by a continuous footpoint driver with an amplitude of 2km/s. We take the transverse temperature and density profile of the initial configuration as in Eq.~\ref{eq:density}-\ref{eq:temperature}, with $T_\mathrm{max}=3\mbox{MK}$ and $d\to\infty$. This profile coincides with the profiles in Sect.~\ref{sec:standing}. This profile is in contrast to the models of \citet{karampelas2017}, which considered cool loops in a hot exterior plasma. The footpoint driver is a transverse oscillation in the velocity, of which the period is $P=315\mathrm{s}$ (which is the kink period). The driver has a uniform velocity field inside the loop and in the $x$-direction, and a dipole field outside the loop, which are connected with a $\tanh$ profile \citep[for more details, see][]{karampelas2017}. Snapshots of the simulation have been included in Fig.~\ref{fig:demovertime_kostas}. 

\begin{figure}
        \includegraphics[width=\linewidth]{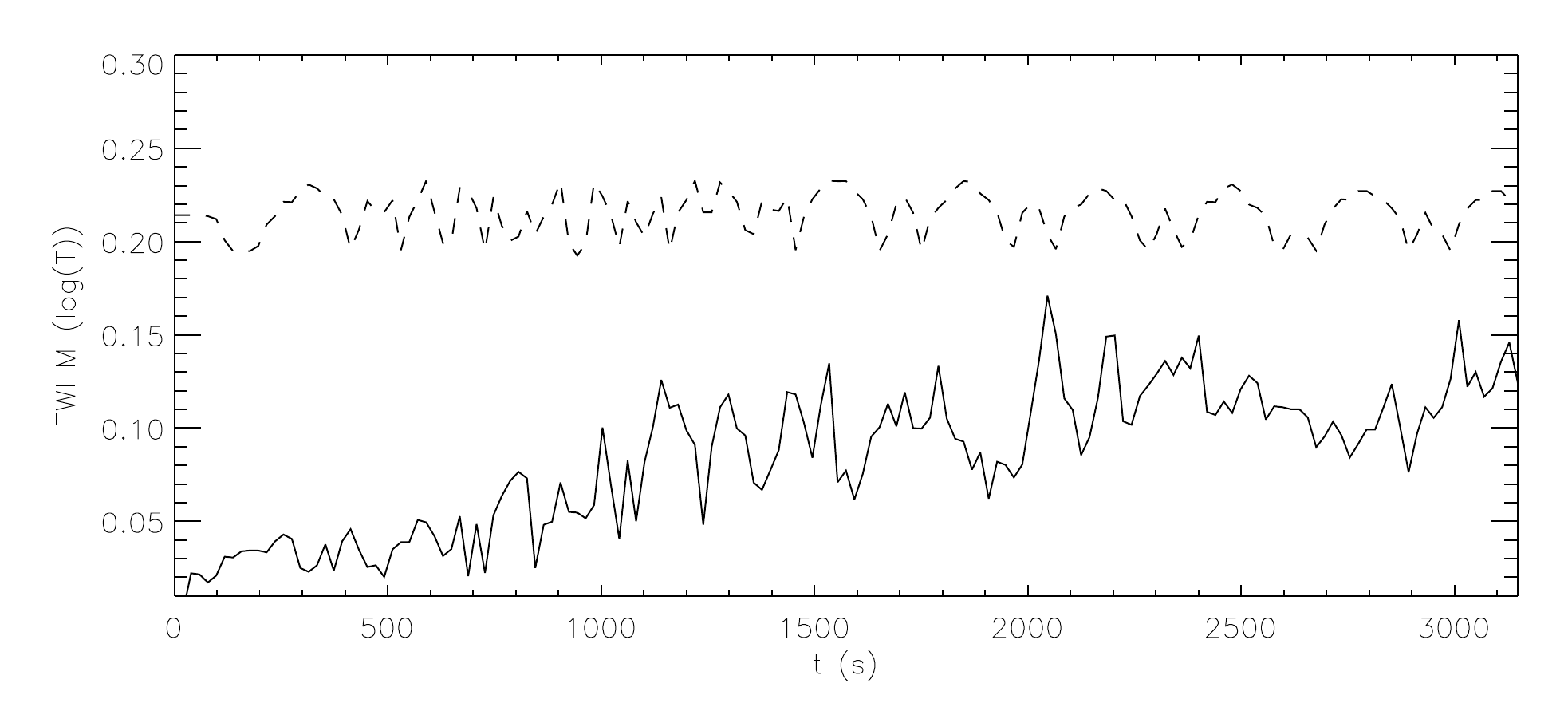}
        \caption{FWHM of the DEM as a function of time, for the driven model. The full line is the FWHM of the numerical DEM, while the dashed line is the FWHM of the DEM from the forward modelling. }
        \label{fig:fwhmdriven}
\end{figure}

Figure~\ref{fig:fwhmdriven} shows the FWHM of the DEM as a function of time. As before, the initial width of the DEM is 0. Over time, the loop becomes more and more turbulent, to finally having completely turbulent cross-sections \citep{karampelas2018}. This has a clear impact on the derived DEM. We see that the FWHM increases steadily. This is surprising, because the kinetic energy in the loop saturates around $t=4-5P=1250-1600\mathrm{s}$. Moreover, the loop becomes fully turbulent after $t=8-9P=2500-2800\mathrm{s}$, and we would have expected that the DEM would also stabilise after that time (by assuming that the plasma is already fully mixed). Some indications are visible in Fig.~\ref{fig:fwhmdriven} that the increase stagnates, but the evidence is insufficient in the current length of simulations. \\
After 10 periods of driving, the width of the DEM has increased to $\log{T}=0.15$, which corresponds to a 41\% increase. Following our results from Sect.~\ref{sec:shell}, the width of the DEM is determined by the initial temperature contrast. The value $\log{T}=0.15$ corresponds well with the values on the right panel of Fig.~\ref{fig:peakfwhm}, given our initial temperature difference is 0.5 (in $\log{T}$). Because of this correspondence, it may indeed be that the FWHM of the DEM would saturate around the end value or slightly higher. Some readers may even suggest that this is shown in Fig.~\ref{fig:fwhmdriven} for $t>2500\mathrm{s}$. It indicates that the plasma mixing by the TWIKH rolls has not been completed yet, because the DEM keeps broadening, even at the final stage of the simulation.  

Here we consider the DEM from the forward modelling. As in Sect.~\ref{sec:shell}, we performed forward modelling of the simulation snapshot and then perform a DEM inversion. However, as in Sect.~\ref{sec:shell}, the DEM from the forward modelling does not reflect the simulated DEM very well. The FWHM of the reconstructed DEM as a function of time is plotted with the dashed line in Fig.~\ref{fig:fwhmdriven}. To compute the FWHM, we have selected a range for fitting of 0.16 (in $\log{T}$) around the peak temperature of the DEM.\\
The reconstructed DEM has a broad peak near the peak temperature of the plasma, perhaps caused by the lack of DEM resolution. This is clearly reflected in the graph for its FWHM: its value is significantly larger than the one of the numerical DEM without forward modelling. The FWHM of the reconstructed DEM seems to be a poor measure for the real FWHM of the numerical DEM. Its behaviour does not show the increase of the DEM width. We conclude that the broadening of the DEM due to loop turbulence by footpoint driving can probably not be observed with imaging telescopes with a limited number of filters, such as AIA. 

\section{Conclusions and discussion}
\label{sec:conclusions}

From the models in the previous sections, we can conclude that the loop turbulence caused by transverse waves results in broadening of the DEM. From multi-shelled, toy models of loops, we found that the width of the DEM is mainly determined by the initial temperature difference between the interior and exterior of the loop.\\ 
For an impulsively excited loop, the width of the DEM saturates after some periods, because the transverse wave energy is completely absorbed in the TWIKH rolls, inhibiting further mixing of the interior, hot plasma with the exterior, cool plasma. The saturation width of the DEM is 0.07, which is significantly lower than multi-thermal loops \citep[e.g.][]{schmelz2014}. Thus, such an impulsively excited loop cannot explain the observed broad DEMs, unless much larger temperature contrasts are considered. \\ 
In our long simulation of driven transverse waves, we find that the width of the DEM increased steadily over the time of the simulation. The final value of the width of 0.15 becomes nearly comparable to some observational values \citep{schmelz2014}. In accordance with our first result, we attribute this width to the initial temperature contrast of the loop with the exterior plasma. With these values of the width of the DEM, it seems that these models of driven transversally oscillating loops can potentially explain the broad, observed DEMs.\\ 
However, our results from forward modelling and DEM reconstruction showed that the increase of the DEM width is not observable with the AIA instrument. This is caused by a bad reconstruction of the peculiar input DEMs, resulting in large overestimates of the FWHM from the forward modelling. \par

Our results have an important implication. In previous work, multi-thermal loops were modelled as being multi-stranded \citep[e.g.][]{bradshaw2011,brooks2012,regnier2014}, and each of the strands had a single temperature at a given time (i.e., a delta-function DEM). Even if we do not take into account the destruction of the strands by the mixing of the transverse waves \citep{magyar2016}, we show here that the omnipresent transverse waves broaden the DEM of a strand beyond a delta-function. The finite width of each strand's DEM is determined by the initial and evolving temperature and the background. Therefore, the number of strands needed to model the broad, observed DEMs can be drastically reduced. The finite width DEMs of single strands are then easily superposed to form broad, observed DEMs. Perhaps even strongly mixed single loops can explain broad DEMs single-handedly in some cases. \\
But importantly, this physical model also implies a transverse correlation between the strands, since a common physical mechanism (the transverse MHD wave) is responsible for their thermodynamic evolution. This implies that a 1D multi-stranded loop model constructed from thousands of strands, if ever possible, needs to be constructed taking into account a specific transverse correlation and temporal correlation.

The fact that our forward modelling and DEM inversion cannot properly reconstruct the input DEMs begs the question if the observation of broad DEMs is indeed a product of a physical process (such as the one described in this paper), or if it just implies that the temperature structure of the loops is unresolved with the limited number of AIA filters. Moreover, one may argue that if the increase of the DEM width by the TWIKH rolls cannot be observed with current imaging instruments, then the observed broadening must be caused by another mechanism. However, inclusion of multiple strands (perhaps two or three) in our model with each different peak temperatures would result in much broader final DEMs, which can potentially be observed by SDO and explain the observed DEMs.\\ On the other hand, spectroscopic instruments (such as Hinode/EIS) have a much finer `temperature resolution' and can potentially study the DEM distribution with sufficient accuracy to follow its evolution over time. Thus, to observationally test the predicted DEM broadening in this paper, spectroscopy in multiple spectral lines with broad temperature coverage is definitely needed. 

The key conclusion of this paper is that the turbulent loop models (which take into account the observed dynamic behaviour of the loops) can explain the observed broad DEMs in the solar corona, equally well as multi-stranded loops. This is an important point, because it means that the turbulent loop model with TWIKH rolls can explain several observed features: (i) small scale transverse motions \citep{antolin2016}, (ii) apparent strands \citep{antolin2014}, (iii) phase shift between Doppler shifts and intensity \citep{antolin2015}, (iv) unchanged filling factors \citep{magyar2016}, (v) unchanged emission measures \citep{magyar2016}, and (vi) broad DEMs (this manuscript). Adding to this list, and generating observables (predicted broadening of DEM), is important to verify this turbulent loop model or to disqualify it.

\begin{acknowledgements}
TVD was supported by an Odysseus grant of the FWO Vlaanderen, the IAP P7/08 CHARM (Belspo), the GOA-2015-014 (KU~Leuven) and the European Research Council (ERC) under the European Union's Horizon 2020 research and innovation programme (grant agreement No 724326). The results were inspired by discussions at the ISSI Bern and at ISSI Beijing. TVD is grateful for the support from the NAOJ Visiting Fellows Program. PA has received funding from the UK Science and Technology Facilities Council (Consolidated Grant ST/K000950/1), the European Union Horizon 2020 research and innovation programme (grant agreement No. 647214) and his STFC Ernest Rutherford Fellowship (grant agreement No. ST/R004285/1). Numerical computations were carried out on Cray XC30 at the Center for Computational Astrophysics, NAOJ.
\end{acknowledgements}

\bibliographystyle{aa}
\bibliography{refs,../refs_patrick}

\end{document}